\documentclass[twocolumn,prl,showpacs]{revtex4}
\usepackage{amstext}
\usepackage{amsmath}            %serve per le subequazioni
\usepackage{amssymb}            %serve per il simbolo "marchio registrato", \circledR
\usepackage{graphicx}           %serve per le figure eps, ps etcyy
\usepackage{latexsym}

\newcommand{\one}{\mbox{$1 \hspace{-1.0mm}  {\bf l}$}}

\begin{document}
\title{Coherent quantum evolution via reservoir driven holonomy}
\author{ Angelo Carollo$^{\diamond}$,
Marcelo Fran\c{c}a Santos$^{\dag}$ and Vlatko Vedral$^{\ddag}$}
\address{$^{\diamond}$DAMTP, University of Cambridge,
Wilberforce Road, Cambridge CB3 0WA, United Kingdom  \\
$^\ddag$School of Physics \& Astronomy, University of Leeds, LS2 9JT, United Kingdom\\
$^\dag$Departamento de F\'{\i}sica, Universidade Federal de Minas
Gerais, Belo Horizonte, 30161-970, MG, Brazil}

\begin{abstract}
We show that in the limit of strongly interacting environment a
system initially prepared in a decoherence-free subspace (DFS)
coherently evolves in time, adiabatically following the changes of the DFS.
If the reservoir cyclicly evolves in time, the DFS states acquire an holonomy.
\end{abstract}

\pacs{03.65.-w 03.65.Vf 03.65.Yz}   %actually, we didn't put the pacs number in the text submitted.
                                    %These are the pacs number that we have specified in the submission form.
 \maketitle

Geometric phases and Holonomies are generated by the presence of a
curvature (the Berry curvature) in the Hilbert space. The natural
way to gain information on this curvature is to perform
interferometry on a state vector which has been dragged along closed
loops in the Hilbert space. To achieve this task the methods
exploited are usually divided into two main categories: in the first
one, a state is driven around the space by means of a {\em coherent}
time evolution~\cite{Berry84}. For example, this can be obtained by
an adiabatic and cyclic motion, which forces the state to explore
eigenspaces of a parameterized Hamiltonian. The second category
consists of all those motions that result from \emph{non-trace
preserving} quantum operations. For example, a series of von Neumann
measurements can be used to evolve wave
functions by projecting the state vector onto a sequence of
overlapping subspaces. As an effect of this
\emph{measurement-induced} motion, a quantum system may indeed
acquire a geometric (Abelian or not)
phase~\cite{AharonovV80}. The
Pancharatnam geometric phase is a typical example of this
category~\cite{AharonovV80,Pancharatnam56}.

Even irreversible quantum processes
such as systems interacting with Markovian reservoirs can be used
to generate geometric evolutions. Many environmental models have the
characteristic of not affecting some particular quantum states when
these are lying in suitably ``protected'' subspaces, so-called
decoherence free subspaces
(DFS)~\cite{PalmaSE96}. States lying in
these subspaces are \emph{stationary}, i.e. they do not evolve in
time. Having control over the reservoir (e.g. through engineered
reservoirs) may imply an indirect control on the protected states of
the system~\cite{CarvalhoMMD01}. In particular, modifications to the
parameters of the reservoir may result in a controlled time
evolution of the protected subspace as a whole. If this motion is
accomplished in a sufficiently smooth way, it can be shown that
states lying in this subspace evolve coherently, thereby acquiring
information about the geometry of the space explored. 

In this letter we will explore the possibility of generating
(Abelian and non-Abelian) holonomies due to a smooth motion of a
DFS in the Hilbert space. It is worth stressing that the existence of a time-dependent DFS is by no means trivial, and its presence often reflects a symmetry preserving evolution.
Under a suitable "adiabatic condition"~\cite{LidarOpen}, it can be shown that a state
lying in a DFS remains inside the subspace and, hence, is rigidly
transported around the Hilbert space together with the DFS.
\emph{The evolution experienced is, in fact, coherent,
although entirely controlled via an incoherent phenomenon}.
When the DFS is eventually brought back to its initial configuration,
\emph{the net effect is an holonomic transformation on the states lying in this subspace}.
Notice that the evolution obtained is \emph{purely geometrical}:
no dynamics affects the system inside the DFS, hence, no dynamical
phase is accumulated during the evolution.

Moreover, we will show that in the first order correction to the
``adiabatic'' evolution the DFS is affected by a decoherence process. 
This effect is unavoidable, at least conceptually, as
far as a non-trivial holonomy is to be obtained.  On the other hand, the decoherence time inside the
DFS can be made \emph{arbitrarily large}, in the limit of
\emph{strong decoherence} and/or slow evolution of the
subspace. In fact, the decoherence affecting the DFS becomes weaker
as the decoherence acting on the outside world gets stronger. This
counter-intuitive effect, is the essence of the adiabatic
approximation. It is, in fact, the fast evolving decoherence process
affecting the outside world which decouples the latter from the DFS.
Alternatively, this evolution can be understood in terms of a ``Zeno
effect,''~\cite{MisraS77} where the action of strong environment can be regarded as
a measuring apparatus continuously monitoring the DFS of the system.
By using this effect in degenerate subspaces it is possible to
realize robust holonomic gates for quantum computation, analogous to
those realized by means of adiabatic
evolutions~\cite{UnanyanSB99}.

Let us first consider a system described by the density operator
$\rho$ evolving under the effect of a Markovian environment. The
decoherence process due to the interaction with such an environment
is described by the following master equation ($\hbar=1$):
\begin{equation}
  \label{eq:mastereq}
  \frac{d\rho}{dt}=-i[H,\rho]-\sum_{k=1}^{n}
  \{\Gamma_k^\dagger \Gamma_k\rho+\rho \Gamma_k^\dagger \Gamma_k\ - 2\Gamma_k
  \rho\Gamma_k^\dagger\},
\end{equation}
where the commutator generates the coherent part of the evolution
and the remaining part represents the effect of the reservoir on the
dynamics of the system. The action of each $\Gamma_k$ (the Lindblad
operators) amounts for the different decohering processes that can
affect the system. In the Markovian formalism, a decoherence free
subspace (DFS) is defined as the common eigenspace of all the Lindblad operators:
%\begin{equation}
%\label{ DFS}
$\text{Span} \{|\psi\rangle \mid \Gamma_k|\psi\rangle
= c_k|\psi\rangle, \forall k \}$. 
%$\text{Span}\{\psi | \Gamma_k\ket{\psi}=c_{k}\ket{\psi}, \forall k \}$
%\end{equation}

Suppose that the evolution is solely due to the action of the environment,
i.e. the presence of a Hamiltonian evolution can be neglected compared
to the effect of the environment ($H=0$).
%Let's also assume for simplicity that the noise model is such that only one Lindblad operator  $\Gamma$ appears in the master equation.
%(the results in this paper can be easily generalized to more general cases).
Consider, now the situation in which the environment depends on some
time-dependent external parameter. This time dependence might be regarded as
either the action of an experimenter who can control some degree of freedom of
the reservoir, or ``drift" of some parameter intrinsically occurring in the
environment. In general, we will assume the rate of this evolution very small
compared to characteristic time scales of the system. Under this assumption,
the evolution is governed by a time-dependent master equation of the form:
\begin{equation}
  \label{eq:mastereq1}
  \frac{d\rho}{dt}=-\sum_k
\{\Gamma_k^\dagger (t)\Gamma_k (t)\rho+\rho \Gamma_k^\dagger (t)\Gamma_k (t)\! - 2\Gamma_k (t)
  \rho\Gamma_k^\dagger (t)\},
\end{equation}
where the operator $\Gamma_k(t)$ are time dependent. For each
instant of time we can consider the eigespaces
$\mathcal{K}(t)=\text{Span} \{|\psi\rangle \mid \Gamma_k(t)|\psi\rangle
= c_k|\psi\rangle, \forall k \}$
 of the instantaneous operators $\Gamma_k (t)$. Let $\Pi (t)$ be
the projector onto $\mathcal{K}(t)$, and $\Pi_{\bot}=\one-\Pi(t)$
the projector onto the orthogonal complement
$\mathcal{K}_{\bot}(t)$. 
Consider the operator 
\begin{equation}D(t)=\sum_k\left(\Gamma_k(t)^\dag \Gamma_k(t)-2c_{k}^{*}(t)\Gamma_{k}(t)+|c_{k}(t)|^{2}\one\right)\text{,}\end{equation} whose Hermitian part $P(t)=(D(t)+D^{\dag}(t))/2$ is positive semidefinite. The space $\mathcal{K}(t)$ can be defined as the kernel of $D(t)$, i.e. $D(t)\Pi(t)=0$.
%the number $n$ of zero eigenvalues of $\Xi (t)$ is the dimension of
%the kernel of  $\Gamma (t)$, and the rank of $\Pi(t)$ (for
%simplicity, let's consider a finite dimensional system).
The assumptions that we will consider are the analogous to the hypotesis of the usual adiabatic theorem~\cite{Messiah,Nenciu80}: (i) $\Pi (t)$ is
a chain of projection operators, smoothly depending on time, (more precisely $\Pi(t)$ norm-twice differenciable~\cite{Nenciu80}) ;
(ii) the non-zero eigenvalues of $P(t)$
are bounded from below by a time independent value $\gamma>0$.

Under these assumptions, we will show that: (a) it
is possible to formulate an adiabatic limit, for which
$\mathcal{K}$ is decoupled from the environmental action, and therefore
it is \emph{decoherence free}; (b) in the adiabatic limit, if the
subspace $\mathcal{K}(t_1)$ at some time $t_1=t_0+T$ coincides with
$\mathcal{K}(t_0)$ at the initial time $t_0$, the neat evolution experienced
by a state belonging to $\mathcal{K}$ is \emph{purely geometrical},
i.e. the evolution between $t_0$ and $t_1$ is described by the \emph{holonomy}
(Abelian or not) associated with the \emph{path} traversed by $\mathcal{K}(t)$
in the Hilbert space; (c) the first order correction to the adiabatic
limit is a source of decoherence, whose time scale is inversely proportional
to the smallest time scale of the decoherence affecting the $\mathcal{K}_\bot$,
the latter being the orthogonal complement of $\mathcal{K}$.

We would like to describe the evolution at some time $t_1=t_0+T$ of states
initially prepared inside the subspace $\mathcal{K} (t_0)$, in the limit
in which the time rate of motion of $\mathcal{K}(t)$ is much smaller than
the smallest characteristic time scale of the system, which is $\gamma^{-1}$.
It is then convenient to introduce a time independent
parameter $s=(t-t_0)/T$, with $s\in [0,1]$ and solve the master equation
in the limit of $T\to \infty$. By following a similar approach of the standard
adiabatic theorem~\cite{Messiah}, let's consider a unitary operator $O(s)$ for which:
%\begin{equation}\label{operatorO}
$\frac{d}{ds}\left(O^{\dagger} \Pi O\right)=0$ with $O(0)=\one$.
%\end{equation}
The operator fulfilling this condition is defined by the equation
$i\frac{d}{ds}O(s)=G(s)O(s)$ (with initial condition
$O^{\dagger}(0)=\one$)~\cite{Messiah, Nenciu80} where the generator is
$G=i\left[\frac{d\Pi}{ds},\Pi\right]+Q(s)$, and $Q(s)$ is an
arbitrary block diagonal Hermitian operator, i.e. $\Pi
Q(s)\Pi_{\bot}=0$. By definition, $O^{\dag}$ describes the change of
picture to the "rotating frame", which rigidly follows the subspace
$\mathcal{K}(s)$. In fact, in this picture, the projector
$\bar{\Pi}(s)\equiv
O^{\dagger}(s)\Pi(s)O^{\dagger}(s)^{\dag}=\Pi(0)$, and the
corresponding subspace $\bar{K}$ is time independent. The freedom in
the choice of the generator $G(s)$ corresponds to the arbitrary
choice of a smooth chain of basis within $\mathcal{K}$ and
$\mathcal{K}_{\bot}$, and it is often referred in this context as
\emph{gauge freedom}. Under re-parametrization $t\to s=(t-t_0)/T$,
and change of frame $\rho\to\bar{\rho}=O^{\dagger}\rho O$,
equation~(\ref{eq:mastereq1}) takes the form:
\begin{equation}
  \label{eq:mastereq2}
  \frac{d\bar{\rho}}{ds} =i[\bar{G},\bar{\rho}]-\gamma T\{\bar{D}(s)\bar{\rho}+\bar{\rho} \bar{D^{\dag}}(s)
  - 2\sum_{k}\bar{\Gamma}_{k}\bar{\rho}\bar{\Gamma}_{k}^{\dag}\}
\end{equation}
with
$\bar{G}=O^{\dag}G O$,
%$\bar{G}=iO^{\dag}\frac{dO}{ds}$
$\bar{\Gamma}_{k}=\left(O^{\dagger}\Gamma_{k}O-c_{k}\one\right)/\sqrt{\gamma}$, and
$\bar{D}=O^{\dagger}DO/\gamma$. Notice that $\bar{\Gamma}_{k}\Pi=0$ and that the operator $\bar{D}$ has been re-normalized, so that its Hermitian part $\bar{P}=(\bar{D}+\bar{D}^{\dag})/2=\sum_{k}\bar{\Gamma}_{k}^{\dag}\bar{\Gamma}_{k}$ has minimum no-null eigenvalue not smaller than $1$.

From equation~(\ref{eq:mastereq2}) it is
possible to see that the incoherent part of the evolution affects only
system states lying outside the subspace $\bar{\mathcal{K}}$.
However, this does not prevent a state initially prepared in
$\bar{\mathcal{K}}$ to be exposed to decoherence. Indeed, the off
diagonal terms $\bar{G}_{\text{off}}\equiv \bar{\Pi}_{\bot} \bar{G}
\bar{\Pi}$ and $\bar{G}_{\text{off}}^{\dagger}$ of $\bar{G}$ can
couple the subspace $\bar{\mathcal{K}}$ to its complementary, and
may eventually spoil the coherence of the former. In order to unwind
the effect of the perturbation introduced by $G$ from the main
decoherence process, it is convenient to introduce a transformation
which allows the evolution to be described in terms of two effectively
decoupled manifolds. To this end, consider the following effective
non-Hermitian Hamiltonian:
\begin{equation}
\tilde{H}(s)\equiv -e^{iS(s)}\left(\eta \bar{G}(s) +i \bar{D}(s) \right)e^{-iS(s)}
\end{equation}
where $\eta\equiv 1/\gamma T$ is the \emph{adiabatic parameter} and
$S(s)$ is a (non-Hermitian) operator defined by the condition that:
($\alpha$)
$\bar{\Pi}\tilde{H}\bar{\Pi}_{\bot}=\bar{\Pi}_{\bot}\tilde{H}\bar{\Pi}=0$
(with $\bar{\Pi}_{\bot}=\one-\bar{\Pi}$) and ($\beta$)
$\bar{\Pi}S\bar{\Pi}=\bar{\Pi}_{\bot}S\bar{\Pi}_{\bot}=0$~\cite{Apostol74}. Let's now
analyze the time evolution of the density matrix under the
transformation $e^{iS}$. To this end consider the operator
$\tilde{\rho}=e^{iS}\bar{\rho}e^{-iS^{\dag}}$. Notice that although
this transformation is not unitary, $\tilde{\rho}$ is still a
positive semi-definite operator, and hence it is a valid density
matrix, up to a normalization factor $N=Tr(\bar{\rho}
e^{i(S-S^{\dag})})> 0$. By assuming for simplicity that the time
derivative of $S$ is negligible, the evolution of $\tilde{\rho}$ is
given by:
\begin{equation}\label{Meqintermediate}
\frac{d\tilde{\rho}}{ds}\simeq -\frac{i}{\eta}[\tilde{H}\tilde{\rho}-\tilde{\rho}\tilde{H}^{\dag}]
+\frac{2}{\eta}\sum_{k}\tilde{\Gamma}_{k}\tilde{\rho}\tilde{\Gamma}_{k}^{\dag},
\end{equation}
with $\tilde{\Gamma}_{k}\equiv e^{iS}\bar{\Gamma}_{k}e^{-iS}$. The advantage of
this expression is that $\tilde{H}$ is now block diagonal and the main coupling
effects are only due to the second term.
We are interested in the evolution of the system in the limit of
$\eta\ll 1$. It is, then, convenient to consider the expansion of $S$ in
series of $\eta$: $S=\eta S_{1}+\eta^{2}S_{2}+O(\eta^{3})$, and, correspondingly,
the expansion of the effective Hamiltonian $\tilde{H}=\tilde{H}_{0}+\eta\tilde{H}_{1}
+\eta^{2}\tilde{H}_{2}+O(\eta^{3})$.
By using the above conditions ($\alpha$) and ($\beta$) it is possible to show that:
\begin{eqnarray}
\tilde{H}_{0}=&-i\bar{D}(s)\text{,}\\
\tilde{H}_{1}=&-\bar{\Pi}\bar{G}\bar{\Pi}
-\bar{\Pi}_{\bot}\bar{G}\bar{\Pi}_{\bot}\text{,}\\
\tilde{H}_{2}=&-\frac{1}{2}\bar{\Pi}[iS_{1},\bar{G}]\bar{\Pi}
-\frac{1}{2}\bar{\Pi}_{\bot}[iS_{1},\bar{G}]\bar{\Pi}_{\bot}\text{,}\label{Htilde2}
\end{eqnarray}
where $S_{1}=\bar{D}^{-1} \bar{G}_{\text{off}}-H.c.$. Notice that $\bar{D}^{-1}$ in the last expression
is well defined, as it is restricted to the subspace $\mathcal{K}_{\bot}$ where $D$ is invertible.
In the limit of $\eta\ll 1$ the evolution of $\tilde{\rho}$ can be expressed as follows:
\begin{equation}\label{LindExp}
\frac{d\tilde{\rho}}{ds}\simeq\frac{1}{\eta}\mathcal{L}_{-1}[\tilde{\rho}]
+\mathcal{L}_{0}[\tilde{\rho}]+\eta\mathcal{L}_{1}[\tilde{\rho}]+O(\eta^{2}).
\end{equation}
The largest contribution in~(\ref{LindExp}) is given by
\begin{equation}\label{mastereqNew-1}
    \mathcal{L}_{-1}[\tilde{\rho}]=-\sum_{k}
  \{\bar{\Gamma}_k^\dagger \bar{\Gamma}_k\tilde{\rho}
  +\tilde{\rho}\bar{\Gamma}_k^\dagger \bar{\Gamma}_k - 2\bar{\Gamma}_k \tilde{\rho}\bar{\Gamma}_k^\dagger\},
\end{equation}
where we also expressed $\tilde{\Gamma}_{k}=\bar{\Gamma}_{k}
+i\eta[S_{1},\bar{\Gamma}_{k}]+O(\eta^{2})$ and retained only the zero-th contributions in $\eta$. By definition of $\bar{\Gamma}_{k}$
eq.~(\ref{mastereqNew-1}) acts trivially on the partial density operator
$\bar{\rho}_{DF}=\bar{\Pi}\tilde{\rho}\bar{\Pi}$, i.e. $\mathcal{L}_{-1}[\bar{\rho}_{DF}]=0$.
Therefore, the most relevant non-trivial evolution affecting the state $\bar{\rho}_{DF}$ is given by the term $\mathcal{L}_{0}$ in the adiabatic expansion:
\begin{equation}\label{MEqZeroOrd}
\mathcal{L}_{0}[\bar{\rho}_{DF}]=i[\bar{G}_{DF},\bar{\rho}_{DF}] \qquad \text{with } \bar{G}_{DF}=\bar{\Pi}\bar{G}\bar{\Pi},
\end{equation}
where the fact that $\bar{\Gamma}_{k}\bar{\rho}_{DF}=0$ has been used.
%The evolution of the $\rho$ up to the first order in $(\gamma T)^{-1}$ is then given by:
%\begin{equation}
%\label{eq:masterapprox}
% \frac{d\bar{\rho}_{DF}}{ds} \simeq-i[G_{DF},\bar{\rho}_{DF}]-\frac{1}{2\gamma T}\{\Lambda\bar{\rho}_{DF}+\bar{\rho}_{DF}\Lambda\}.
%\end{equation}
%where $G_{DF}=\Pi_0 G\Pi_0$ and
%\begin{equation}
%\label{eq:defLambda}
% \Lambda=G_{01}\bar{D}(s)^{-1}G_{01}^\dag,
%\end{equation}
%with $G_{01}=\Pi_0 G (\one-\Pi_0)$. Notice that, although $D(t)$ is not invertible, $(\one-\Pi_0) D^{-1}(s) (\one-\Pi_0)$ is well defined.
This expression explicitly demonstrates that retaining only the
terms up to the zeroth order in the adiabatic parameter $\eta$
yields to an evolution for $\bar{\rho}_{DF}$ which is \emph{unitary}, and,
therefore, \emph{coherent} and \emph{trace preserving}. This shows
the existence of an \emph{adiabatic limit} for $\eta\to 0$, in which
the evolution is confined - by the decoherence process itself - into
a suitable subspace, in which the evolution \emph{maintains its
coherence}. It is worth noticing that, due to the non-unitary
transformation $e^{iS}$, the new decoherence-free subspace
$\check{\mathcal{K}}$ is slightly modified from the original
$\bar{\mathcal{K}}$. Indeed, the presence of the generator of the
adiabatic motion, $G$, deformes the DFS, whose (generalized)
projectors are to be identified with
%\begin{equation}\label{newprojectors}
$\bar{\Pi} \to \check{\Pi}=e^{iS}\bar{\Pi}e^{-iS^{\dagger}}\simeq \bar{\Pi}+i\eta\{S_{1},\bar{\Pi}\}\text{.}$
%\end{equation}
As expected, in the limit $\eta \to 0$, $\check{\mathcal{K}}$ converges to $\bar{\mathcal{K}}$.

The evolution of the partial density matrix $\bar{\rho}_{DF}$ can be easily
solved in the adiabatic limit and formally expressed as:
\begin{equation*}
\label{eq:formalsol}
\bar{\rho}=U(s)\rho(0)U^\dag(s) \text{ with }U(s)=\mathcal{P} \exp \left(i\int_0^s \bar{G}_{DF}(\tau) d\tau\right),
\end{equation*}
where $\mathcal{P}$ is a path ordering operator.
Let's now consider a closed evolution of $\mathcal{K}$, i.e. an evolution for which $\Pi(t_{1})=\Pi(t_{0})$.
Then it is possible to express the total evolution as:
\begin{equation*}
\label{eq:holonomy}
U(1)=\mathcal{P} \exp \left(i\int_0^1 \bar{G}_{DF}(\tau) d\tau\right)
=\mathcal{P} \exp \left(\oint \vec{A} d\vec{\lambda} \right),
\end{equation*}
where, the right hand side expresses the time evolution in terms of a path integral.
The operator $\vec{A}$, is a (non-Abelian) \emph{holonomic connection}, defined as a vector of components
\begin{equation}
\label{eq:Connection}
A_{\lambda_i}= -\bar{\Pi} O^{\dag} \frac{\partial O(\lambda_1,\lambda_2 \dots ) }{\partial\lambda_{i}} \bar{\Pi},
\end{equation}
where $\lambda's$ are a set of variables parameterizing the
transformations $O(s)$. $U(1)$ has an inherently geometrical nature:
it is independent of the time rate and only a function of the
structure of the underlying Hilbert space. The non-trivial value of
$U(1)$ is in fact a manifestation of the curvature of the Hilbert
space experienced by $\tilde{\rho}$ when it is dragged along in the
subspace $\check{\mathcal{K}}$. The connection $\vec{A}$ behaves as
a proper gauge potential: under a change of basis in $\mathcal{K}$,
$O(s)\to O^{\prime}(s)=O(s)\Omega (s)$, where
$[\bar{\Pi},\Omega]=0$, $\vec{A}$ transforms as
$\vec{A}\to\vec{A}^{\prime}=\Omega^{-1} \vec{A}
\Omega+\Omega^{-1}\nabla\Omega$. By choosing a convenient gauge for
which $\Pi O^{\prime}(1)\Pi=\Pi$, it is possible to show that the
net effect on the state $\rho\simeq\tilde{\rho}_{DF}$ after a complete cyclic
evolution is given by:
\begin{equation}\label{Invariant}
U^{\prime}(1)=\mathcal{P}\exp \oint i\left[\frac{d\Pi}{ds},\Pi\right] ds\text{,}
\end{equation}
which is independent of the gauge chosen~\cite{Zanardi01}.
Equation~(\ref{MEqZeroOrd}) implies that the largest relevant
non-coherent contribution to the dynamics of $\tilde{\rho}_{DF}$ can
originate only at a further order in the adiabatic parameter $\eta$, i.e. from $\mathcal{L}_{-1}$.
By substituting equation~(\ref{Htilde2}) in~(\ref{Meqintermediate}),
and by making again use of the fact that
$\bar{\Gamma}_{k}\bar{\rho}_{DF}=0$, it is possible to show that the first
order correction to an evolution restricted to $\check{\mathcal{K}}$
can be given in the following closed form:
\begin{eqnarray}\label{mastereqNew1}
    \mathcal{L}_{1}[\bar{\rho}_{DF}]&=&-i[Z,\bar{\rho}_{DF}]-\\&-&\sum_{k}
  \{\Lambda_k^\dagger \Lambda_k\bar{\rho}_{DF}+\bar{\rho}_{DF}\Lambda_k^\dagger \Lambda_k
  - 2\Lambda_k \bar{\rho}_{DF}\Lambda_k^\dagger\}\text{,}\nonumber
\end{eqnarray}
where $
%\begin{eqnarray*}\label{mastereqNew1a}
\Lambda_{k}\equiv\bar{\Gamma}_{k}S_{1}=\bar{\Gamma}_{k}\bar{D}^{-1}G_{\text{off}}$ and 
$Z\equiv S_{1}^{\dag}R S_{1}=iG_{\text{off}}^{\dag}\left(\bar{D}^{-1\dag}-\bar{D}^{-1}\right)G_{\text{off}}$.
%\end{eqnarray*}
%and $R=i\sum_{k}\left( c_{k}\bar{\Gamma}_{k}^{\dag}-c_{k}^{*}\bar{\Gamma}_{k} \right)$ is the anti-Hermitian part of $\bar{D}$.
The super-operator $\mathcal{L}_{1}$ generates an incoherent
evolution with a typical time scale on the order of $\eta^{-1}$. It
is worth stressing that, although eq.~(\ref{mastereqNew1}) looks
similar to a master equation in the Lindblad form, it generates an
evolution that, when restricted to the subspace $\check{\mathcal{K}}$, can be non-trace preserving. This is, in fact, the case when
$\Pi_{\bot}\Lambda_{k}\Pi\ne0$. As, by construction, the evolution in the whole Hilbert space is trace-preserving, this clearly indicates a leakage of population from the DFS to the
orthogonal subspace, with a time rate on the order $\eta^{-1}$.
 
Remarkably, since $\mathcal{L}_{1}$ is the
largest order incoherent term affecting the subspace
$\check{\mathcal{K}}$ and it is proportional to $\gamma^{-1}$, it
implies a decoherence time in  $\check{\mathcal{K}}$ which is
inversely proportional to the original decoherence time affecting
the complementary space: a stronger decoherence in
$\check{\mathcal{K}}_{\bot}$ implies weaker environmental effects in
$\check{\mathcal{K}}$. This apparently counter-intuitive phenomenon
is the essence of the adiabatic approximation. It is the fast
dynamics associated with the incoherent processes acting on
$\check{\mathcal{K}}_{\bot}$ which is responsible for decoupling the
subspace $\check{\mathcal{K}}$ from its complement. This fast
evolution averages out the effect of $G_{\text{off}}$ and results in
a decoherence time scale in $\check{\mathcal{K}}$ which is
\emph{quadratic} in $T$. This is the reason why, in spite of the
relatively long time scale $T$ needed for the adiabatic
approximation, the system in $\check{\mathcal{K}}$ is guaranteed to
be coherent for a time which scales with a further order of
magnitude in $T$. Alternatively, the same result can be interpreted
from the perspective of a quantum Zeno effect. The decoupling
between $\check{\mathcal{K}}$ and $\check{\mathcal{K}}_{\bot}$ can
be regarded as the effect of a continuous measurement process
performed by the environment. The latter continuously monitors
whether the system state leaks to $\check{\mathcal{K}}_{\bot}$ and
projects it back into $\bar{\mathcal{K}}$ in a time scale of
$\gamma^{-1}$, which is much faster than the leakage process induced
by $G_{\text{off}}$.

On the light of the previous discussion on $\mathcal{L}_{1}$, a
remark on the generator of the holonomy is in order. Notice that
equation~(\ref{Invariant}) implies that,  after a closed motion, a
non-trivial net evolution is achieved only if
$i[\frac{d\Pi}{ds},\Pi]=G_{\text{off}}+G_{\text{off}}^{\dagger}$ is
non-vanishing.
%Indeed, if $G_{\text{off}}$ is zero for the whole evolution, the connection can be expressed as $A_\lambda=(\partial_{\lambda}V) V^\dag$, where $V(s)=\Pi O(s) O(0)\Pi$. $A$ is, then, a \emph{pure gauge} and, hence, produces a trivial holonomy $U(1)=\Pi O(1)O(0)\Pi=\one_{\bar{\mathcal{K}}}$.
%
Indeed $[\frac{d\Pi}{ds},\Pi]=0$ implies $d\Pi/ds=0$, which yields
to a trivial motion of the subspace $\mathcal{K}$, and the resulting
holonomy has to be trivial. This is in agreement with the idea that the
holonomy is the result of the curvature of the Hilbert space experienced
by the subspace $\mathcal{K}(t)$ as the latter traverses a non-trivial closed motion.
Therefore, for an holonomic evolution in $\bar{\mathcal{K}}$, the latter
subspace needs to be coupled to $\bar{\mathcal{K}}_{\bot}$ via $G_{\text{off}}$,
which, as seen in eq.~(\ref{mastereqNew1}),
exposes $\tilde{\rho}_{0}$ to environmental effects. This implies that
decoherence on $\bar{\mathcal{K}}$ is in principle unavoidable, as far as
a non-trivial holonomy is concerned. On the other hand, as seen before,
$\mathcal{L}_{1}$ appears as a small perturbation in the adiabatic expansion,
and therefore its effects are adiabatically eliminated in the limit of $T$ large enough.

The results presented in this manuscript reinforce the idea that
geometric evolutions not only do not depend on the dynamical
procedures used to generate them, but can even be obtained in the
presence of strong coupling to external reservoirs (non-unitary
evolutions). As shown here, the interaction with the
environment can be used to adiabatically manipulate quantum states in a coherent way, and generate holonomies in arbitrary
dimensional DFS. Apart from their fundamental interest, these results, combined with techniques for reservoir engineering, may provide alternative approach for robust quantum
computation. As for any other geometrical procedure, this holonomic evolution has an intrinsically fault tolerant nature~\cite{DeChiaraP03}, and due its very nature it is free from dynamical contributions, which are known to be generally more sensitive to errors.

We acknowledge the support of EU, under TOPQIP project. MFS acknowledges the support of CNPq. V.V. acknowledges also support from EPSRC and the British Council in Austria.

%\bibliographystyle{unsrt}
%\bibliography{engeres}

\end{document}